\documentclass[sigconf]{acmart}

\usepackage{balance}
\usepackage[capitalize]{cleveref}

\usepackage{algorithm}
\usepackage{algorithmic}

\usepackage{mathtools}
\usepackage{bm} 

\usepackage{multirow}
\usepackage{tabularx}
\usepackage{adjustbox}
\usepackage{colortbl}
\usepackage{arydshln}
\usepackage{makecell}
\usepackage{tablefootnote}

\usepackage[absolute,overlay]{textpos}
\usepackage{subcaption}

\usepackage{enumitem}
\usepackage{xspace}
\usepackage{soul}
\usepackage[most]{tcolorbox}
\usepackage{listings}

\usepackage{pifont}
\usepackage{gensymb}

\usepackage[disable, textwidth=0.95in]{todonotes} 

\newcommand{\ours}{NeocorRAG\xspace}

\definecolor{newblue}{RGB}{215,238,249}
\definecolor{my_green}{RGB}{40,154,121}
\definecolor{my_yellow}{RGB}{255,165,0}
\definecolor{my_red}{RGB}{176,46,46}
\definecolor{darkgreen}{rgb}{0,0.5,0}
\definecolor{wkblue}{RGB}{210, 230, 250}
\definecolor{wkgreen}{RGB}{226,240,217}

\usepackage{tikz} 
\definecolor{OliveGreen}{rgb}{0.33, 0.42, 0.18}

\definecolor{MyOrange}{HTML}{D95F02}
\definecolor{MyPurple}{HTML}{8856A7}  
\newtcolorbox{correctanswer}[1][]{
    enhanced, breakable, colframe=OliveGreen, colback=OliveGreen!10!white, 
    sharp corners, boxsep=0pt, left=5pt, right=5pt, top=6pt, bottom=6pt,
    boxrule=0pt, leftrule=4pt, #1
}
\newtcolorbox{orangeanswer}[1][]{
    enhanced, breakable, colframe=MyOrange, colback=MyOrange!10!white, 
    sharp corners, boxsep=0pt, left=5pt, right=5pt, top=6pt, bottom=6pt,
    boxrule=0pt, leftrule=4pt, #1
}
\newtcolorbox{analyzeanswer}[1][]{
    enhanced, breakable, colframe=MyPurple, colback=MyPurple!10!white, 
    sharp corners, boxsep=0pt, left=5pt, right=5pt, top=6pt, bottom=6pt,
    boxrule=0pt, leftrule=4pt, #1
}

\lstdefinestyle{mystyle}{
    language=Python, escapeinside={(*}{*)}, breaklines=true, breakatwhitespace=true,
    aboveskip=4pt, belowskip=4pt, basicstyle=\small\ttfamily\linespread{0.9},
    commentstyle=\color{black}\rmfamily, keywordstyle=\color{blue}\bfseries,
    stringstyle=\color{purple}, showstringspaces=false
}

\AtBeginDocument{%
  }

\copyrightyear{2026}
\acmYear{2026}
\setcopyright{cc}
\setcctype{by}

\acmConference[WWW '26]{Proceedings of the ACM Web Conference 2026}{April 13--17, 2026}{Dubai, United Arab Emirates.}
\acmBooktitle{Proceedings of the ACM Web Conference 2026 (WWW '26), April 13--17, 2026, Dubai, United Arab Emirates}
\acmISBN{979-8-4007-2307-0/2026/04}
\acmDOI{10.1145/3774904.3792093}

\begin{document}

\title{\ours: Less Irrelevant Information, More Explicit Evidence, and More Effective Recall via Evidence Chains}

\author{Shiyao Peng}
\orcid{0009-0004-4650-0673}
\email{psy200104@bupt.edu.cn}
\affiliation{%
  \institution{Beijing University of Posts and Telecommunications}
  \city{Beijing}
  \country{China}
}

\author{Qianhe Zheng}
\orcid{0009-0005-7404-0157}
\email{zhengqianhe@bupt.edu.cn}
\affiliation{%
  \institution{Beijing University of Posts and Telecommunications}
  \city{Beijing}
  \country{China}
}

\author{Zhuodi Hao}
\orcid{0009-0002-9178-1457}
\email{jodiehao@bupt.edu.cn}
\affiliation{%
  \institution{Beijing University of Posts and Telecommunications}
  \city{Beijing}
  \country{China}
}

\author{Zichen Tang}
\orcid{0000-0002-0244-4970}
\email{TangZichen@bupt.edu.cn}
\affiliation{%
  \institution{Beijing University of Posts and Telecommunications}
  \city{Beijing}
  \country{China}
}

\author{Rongjin Li}
\orcid{0009-0001-7563-592X}
\email{lirongjin@bupt.edu.cn}
\affiliation{%
  \institution{Beijing University of Posts and Telecommunications}
  \city{Beijing}
  \country{China}
}

\author{Qing Huang}
\orcid{0009-0001-6733-019X}
\email{huangqing@bupt.edu.cn}
\affiliation{%
  \institution{Beijing University of Posts and Telecommunications}
  \city{Beijing}
  \country{China}
}

\author{Jiayu Huang}
\orcid{0009-0004-8023-9157}
\email{jyxhuangjiayu@bupt.edu.cn}
\affiliation{%
  \institution{Beijing University of Posts and Telecommunications}
  \city{Beijing}
  \country{China}
}

\author{Jiacheng Liu}
\orcid{0009-0003-6290-7767}
\email{Liujiacheng@bupt.edu.cn}
\affiliation{%
  \institution{Beijing University of Posts and Telecommunications}
  \city{Beijing}
  \country{China}
}

\author{Yifan Zhu}
\orcid{0000-0002-7695-1633}
\email{yifan_zhu@bupt.edu.cn}
\affiliation{%
  \institution{Beijing University of Posts and Telecommunications}
  \city{Beijing}
  \country{China}
}

\author{Haihong E}
\orcid{0000-0003-2087-586X}
\authornote{Corresponding author.}
\email{ehaihong@bupt.edu.cn}
\affiliation{%
  \institution{Beijing University of Posts and Telecommunications}
  \city{Beijing}
  \country{China}
}

\renewcommand{\shortauthors}{Shiyao Peng et al.}

\begin{abstract}

Although precise recall is a core objective in Retrieval-Augmented Generation (RAG), a critical oversight persists in the field: improvements in retrieval performance do not consistently translate to commensurate gains in downstream reasoning. To diagnose this gap, we propose the \textbf{Recall Conversion Rate (RCR)}, a novel evaluation metric to quantify the contribution of retrieval to reasoning accuracy. Our quantitative analysis of mainstream RAG methods reveals that as \texttt{Recall@5} improves, the RCR exhibits a near-linear decay. We identify the neglect of retrieval quality in these methods as the underlying cause. In contrast, approaches that focus solely on quality optimization often suffer from inferior recall performance. Both categories lack a comprehensive understanding of retrieval quality optimization, resulting in a trade-off dilemma. To address these challenges, we propose \textbf{comprehensive retrieval quality optimization criteria} and introduce the \textbf{\ours{}} framework. This framework achieves holistic retrieval quality optimization by systematically mining and utilizing \textbf{Evidence Chains}. Specifically, \ours{} first employs an innovative activated search algorithm to obtain a refined candidate space. Then it ensures precise evidence chain generation through constrained decoding. Finally, the retrieved set of evidence chains guides the retrieval optimization process. Evaluated on benchmarks including HotpotQA, 2WikiMultiHopQA, MuSiQue, and NQ, \ours{} achieves \textbf{SOTA} performance on both 3B and 70B parameter models, while consuming less than 20\% of tokens used by comparable methods. This study presents an efficient, training-free paradigm for RAG enhancement that effectively optimizes retrieval quality while maintaining high recall. Our code is released at \url{https://github.com/BUPT-Reasoning-Lab/NeocorRAG}.

\end{abstract}




\maketitle

\section{Introduction}
\label{sec:introduction}

Retrieval-Augmented Generation (RAG)~\cite{gao2023retrieval,fan2024ragsurvey2,rag-llm-survey} has emerged as a standard, non-parametric approach to provide Large Language Models (LLMs) with up-to-date, factual information, thus mitigating hallucinations ~\cite{lewis2020nlptasks,RAGefficiency,mallen2022not,zhao2024ai-generated}.
\begin{figure}[t]
	\centering
	\includegraphics[width=\columnwidth]{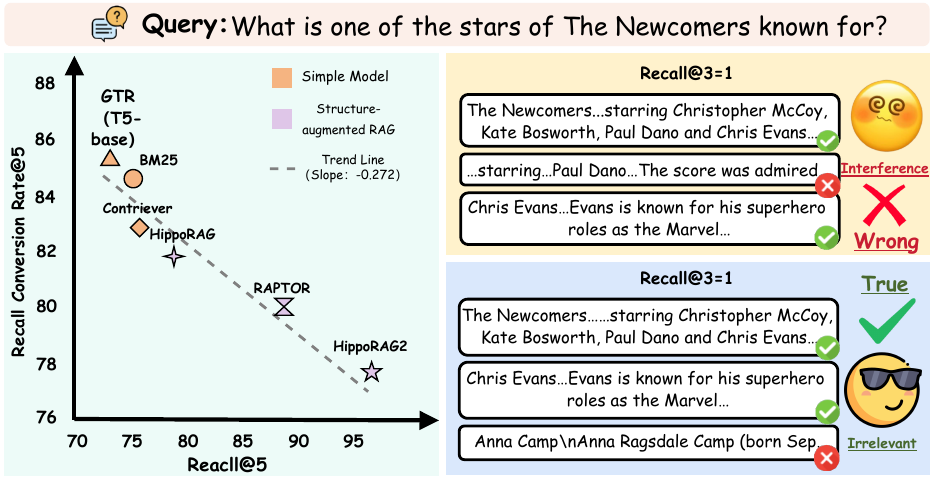} 

   \caption{The inadequacy of recall metrics in capturing retrieval quality misleads RAG optimization and hinders reasoning performance improvement.
    \textbf{(Left)} The increase in the \texttt{Recall@5} score of retrieval methods is accompanied by a decrease in the Recall Conversion Rate (RCR). 
    \textbf{(Right)} For retrieved texts both with a \texttt{Recall@3=1} score, the upper example contains misleading information for the model, while the lower one contains only completely irrelevant information, leading to different QA results. This indicates that \texttt{Recall@n} can only measure ``whether relevant documents are hit'' but fails to accurately assess retrieval quality.}
    \vspace{-1.5em}
	\label{fig:first_example}
\end{figure}

\begin{figure*}[t]
        \centering	
        \includegraphics[width=0.8\textwidth]{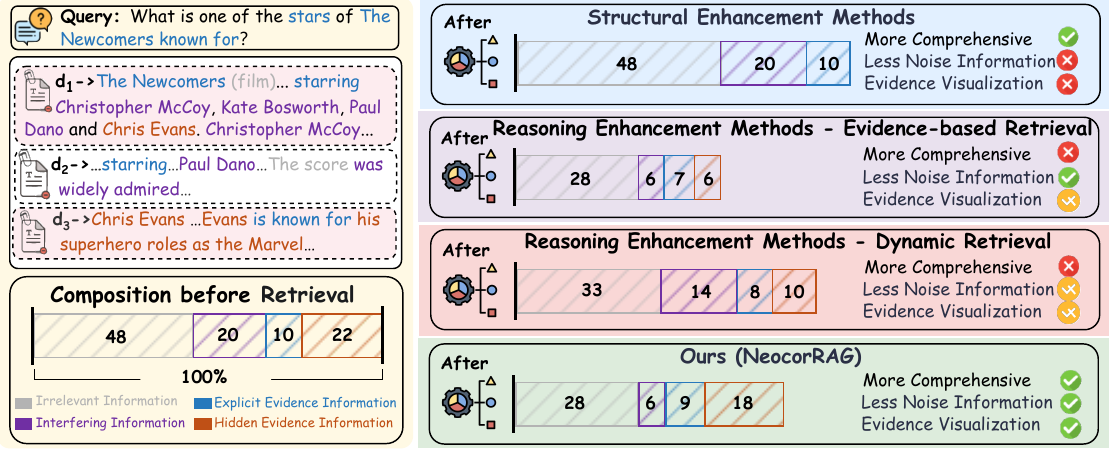}	\caption{Comparison of existing methods based on proposed retrieval quality optimization criteria, evaluating two method categories against three standards (\textit{More Comprehensive}, \textit{Less Noise Information}, \textit{Evidence Visualization}).  Structure Enhancement Methods (top) excel in comprehensiveness but introduces significant noise and lacks evidence visualization. Reasoning Enhancement Methods (middle), which include evidence-based retrieval and dynamic retrieval, show varied improvements in noise suppression and evidence visualization but are deficient in comprehensiveness. In contrast, \ours{} (bottom) simultaneously satisfies all three criteria, achieving a balanced and superior optimization of retrieval quality.}
    \vspace{-1em}
\label{fig:2}	
\end{figure*}
Standard RAG systems~\cite{chen2024dense,contriever,santhanam2022colbertv2,sarthi2024raptor}, which rely on simple vector-based retrieval mechanisms, have demonstrated excellent recall performance in basic Question Answering (QA) tasks~\cite{kwiatkowski2019NQ}. Structure-enhanced RAG methods, such as HippoRAG~\cite{jimenez2024hipporag} and RAPTOR~\cite{sarthi2024raptor}, have further achieved impressive results on more challenging QA benchmarks~\cite{yang2018hotpotqa,2WikiMultiHopQA,trivedi2022musique}. Among them, HippoRAG2~\cite{gutierrez2025hipporag2} stands out by reaching a passage \texttt{Recall@5} of 96.3\% on the HotpotQA benchmark~\cite{yang2018hotpotqa}, seemingly proving that RAG systems can already retrieve relevant content accurately.


However, its corresponding QA F1 score of only 75.5\% presented a stark contrast. This discrepancy suggests that existing retrieval metrics fail to capture how retrieved content is actually utilized during downstream reasoning.  Motivated by this observation, we introduce the \textbf{Recall Conversion Rate (RCR)}, a novel evaluation metric designed to quantify the contribution of retrieved content to reasoning. Based on this evaluation metric, we quantitatively evaluated the main RAG methods. Our analysis reveals that improvements in retrieval recall do not correlate with gains in reasoning accuracy. As shown in the left panel of Figure~\ref{fig:first_example}, the RCR exhibits a near-linear decay as \texttt{Recall@5} increases, exposing a ``high recall, low conversion'' phenomenon. As illustrated in Figure~\ref{fig:first_example}, even with optimal recall scores, the retrieved content can still contain noisy text that severely interferes with the model's reasoning.


This observation exposes a fundamental limitation of commonly used retrieval metrics, such as \texttt{Recall@n}~\cite{recall} and \texttt{NDCG}~\cite{ndcg}, which prioritize surface-level matching of relevant snippets while overlooking whether the retrieved content truly provides faithful, non-misleading evidence that supports downstream reasoning. Consequently, improvements in retrieval metrics often fail to yield corresponding gains in reasoning performance.



Reasoning-enhanced methods have recognized the importance of retrieval quality for reasoning performance, yet they fall into a ``trade-off'' dilemma. \textbf{Evidence-based retrieval methods} attempt to suppress noise by extracting a small set of explicit evidence: they either identify title-anchored evidence via document triples~\citep{fang2024trace} or directly synthesize evidence sentences from external knowledge sources via generative retrieval~\citep{li2024retrollm}. While effective at filtering interference, these approaches often bias the retained context toward shallow, explicitly expressed facts and fail to surface deeper or latent associations within the retrieved texts, which cause some correctly retrieved information to be discarded. \textbf{Dynamic retrieval methods}, such as CoRAG~\citep{wang2024corag}, aim to expand evidence coverage through multi-step retrieval chains but rely on learned retrieval policies that are constrained by the training data, leading to limited generalization and unstable recall behavior across model scales.



In fact, both types of methods lack a systematic analysis of the intrinsic components of retrieval quality. From the perspective of how information supports reasoning, retrieved texts can be classified into four categories: \textbf{Irrelevant Information}, \textbf{Interfering Information}, \textbf{Explicit Evidence}, and \textbf{Hidden Evidence}. Irrelevant information has a negligible impact on the reasoning process; interfering information can mislead the model's judgment; whereas explicit and hidden evidence are the critical pillars for achieving correct reasoning.

Therefore, we propose more systematic criteria for retrieval quality optimization:
\begin{enumerate}[label=(\arabic*), itemsep=0.1em, topsep=0.2em, leftmargin=*]
    \item \textbf{More Comprehensive:} Ensuring broad coverage of the retrieved information.
    \item \textbf{Less Noise Information:} Suppressing the inclusion of interfering noise.
    \item \textbf{Evidence Visualization:} Enhancing the ability to identify and associate hidden evidence.
\end{enumerate}

Using these criteria, we perform a more systematic analysis of the deficiencies in the two main categories of existing methods in Figure~\ref{fig:2}. Neither of these approaches can fully satisfy all the optimization standards. To fully address the challenge of optimizing retrieval quality, we propose \textbf{\ours{}}, a framework that enhances retrieval quality by deeply mining evidence chains within retrieved texts. Evidence chains mining is realized through two key strategies: (1) we design a novel \textbf{activated path search algorithm} that starts from the key entities of the question to identify a candidate evidence space within the subgraph of the texts, thereby thoroughly mining potential evidence chains while substantially excluding irrelevant information; and (2) we then utilize \textbf{constrained decoding} to precisely search for faithful evidence chains within the candidate set. Relying on the resulting set of evidence chains, our approach removes interfering noise and explicates hidden associative evidence, all while ensuring no loss in recall performance.

We validate \ours{} on four widely used QA benchmarks, including NQ~\cite{kwiatkowski2019NQ}, MuSiQue~\citep{trivedi2022musique}, 2WikiMultiHopQA~\citep{2WikiMultiHopQA}, and HotpotQA~\citep{yang2018hotpotqa}. In our experiments, using lightweight models from the Llama 3~\cite{dubey2024llama3herd} and Qwen2.5~\cite{qwen2.5} series for constrained decoding, \ours{} demonstrates consistent state-of-the-art (SOTA) performance across all benchmarks in both 3B (Llama-3.2-3B-Instruct) and 70B (Llama-3.3-70B-Instruct) QA settings, while consuming less than 20\% tokens used by comparable methods. In identical QA prompt settings, compared to baseline before retrieval quality optimization, the F1 score improves by up to \textbf{9.4 percentage points} and the RCR increases by up to \textbf{9.6 percentage points}.

As a universal \textbf{plug-and-play} framework, \ours{} offers a novel perspective on retrieval optimization for RAG systems. In summary, our core contributions are threefold:
\begin{enumerate}[label=(\arabic*), itemsep=0.2em, topsep=0.2em, leftmargin=*]
    \item We reveal the limitations of existing retrieval optimization methods in enhancing retrieval quality, propose more comprehensive criteria for retrieval quality optimization, and introduce the \ours{} framework to effectively address these issues through evidence-chain-driven optimization, providing a clear direction for RAG system enhancement.
    \item We design a novel deep evidence mining method that optimizes retrieval quality by thoroughly mining associative evidence while controlling irrelevant information.
    \item \ours{} achieves SOTA performance on four classic QA benchmarks, being lightweight, efficient, training-free, and plug-and-play.
\end{enumerate}

\section{Related Work}
\label{sec:related_work}

RAG integrates retrieval modules with language models to improve factual accuracy and reduce hallucinations, establishing itself as a core paradigm for knowledge-intensive tasks~\citep{zhao2024ai-generated}. Existing approaches follow two main optimization directions: \textbf{structural enhancement}, incorporating structural information to improve recall, and \textbf{reasoning enhancement}, aligning retrieval with the model’s reasoning needs.

The first category focuses on \textbf{Structural Enhancement}~\citep{KGRAG,wei2025KG2RAG,xu2025DRAG,he2024G-retriever,graph-reranking,Graph-EnhancedRAG}. These methods aim to address the lack of meaning construction~\citep{Klein_Moon_Hoffman_2006} and associative linking~\citep{Suzuki_2005} in traditional RAG recall by leveraging rich structural information. For instance, GraphRAG~\citep{han2024graphrag} manages documents and entities with relations through graph community detection algorithms; HippoRAG~\citep{jimenez2024hipporag} simulates the mechanism of hippocampal memory to perform associative retrieval; and RAPTOR~\citep{sarthi2024raptor} builds a hierarchical summary through recursive embeddings, using a Gaussian Mixture Model (GMM) to detect clusters of documents for summarization. These methods excel in comprehensiveness and achieve high recall. However, they often lack effective noise reduction and do not support evidence visualization, limiting their ability to convert high recall into improved reasoning performance.

The second category focuses on \textbf{Reasoning Enhancement}~\citep{asai2023selfrag, jeong2024adaptive, niu2024judgerank, chirkova2025provence}. These methods concentrate on optimizing retrieval based on the model's reasoning needs. For example, IRCoT~\citep{trivedi2022interleaving} uses Chain-of-Thought prompting to guide alternating retrieval and reasoning steps; CoRAG~\citep{wang2024corag} trains a model for dynamic retrieval queries by designing a retrieval chain data generation strategy; Trace~\citep{fang2024trace} iteratively retrieves reasoning chains from logically connected knowledge triples to answer questions; and RetroLLM~\citep{li2024retrollm} proposes a unified retrieval-generation framework that enables the LLM to directly generate fine-grained evidence from the knowledge base. These methods improve noise reduction and evidence visualization to some extent. However, by focusing on localized reasoning paths, they often compromise completeness and incur significant computational overhead.

In general, existing methods exhibit complementary limitations in the optimization of retrieval quality, which continue to hinder further improvements in downstream reasoning performance. 
More broadly, the challenge of organizing and utilizing retrieved information to better support complex reasoning has long been recognized in information retrieval research, including classical digital library and knowledge organization systems such as Greenstone~\citep{witten2000greenstone} and KEA~\citep{witten1999kea}. These observations motivate the need for more systematic criteria for optimizing retrieval quality and form the basis for the problem setting explored in this work.
\section{Methodology}
\label{sec:Methodology}
\begin{figure*}[t]
        \centering	
        \includegraphics[width=0.95\textwidth]{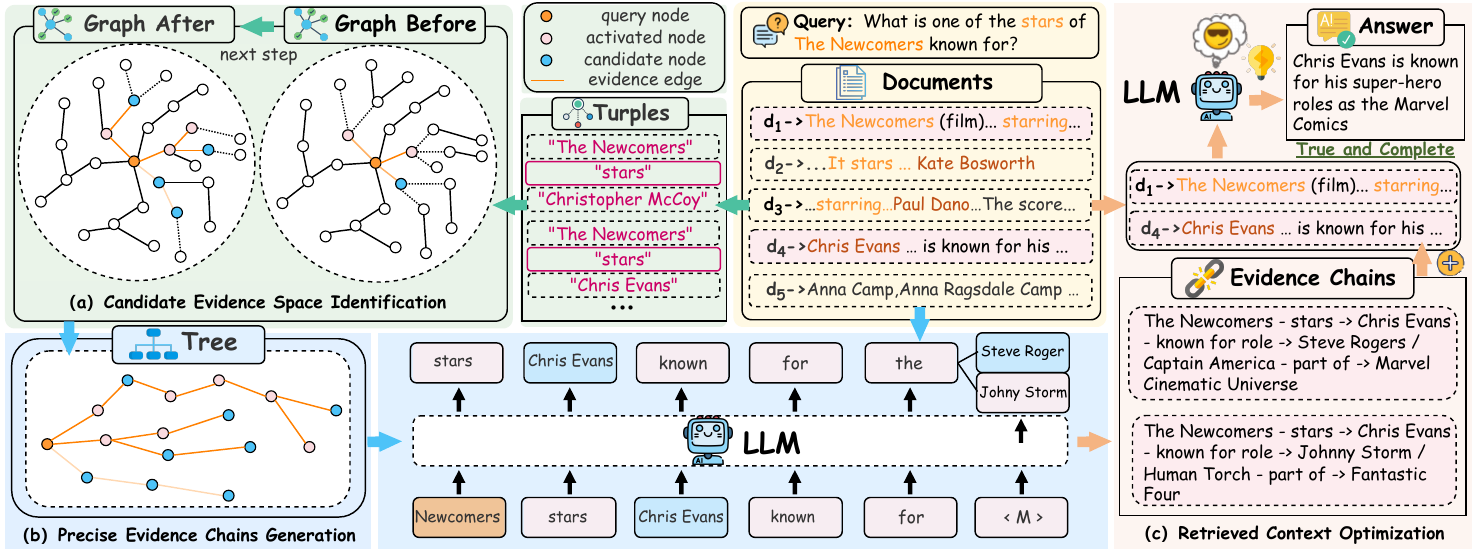}
        
    \caption{Overall framework of \ours{}. Our approach enhances retrieval quality through three sequential stages. 
    \textbf{(a) Candidate Evidence Space Identification:} An activation-based search algorithm dynamically explores paths in the document subgraph to identify candidate evidence chains (orange paths), scoring them with a comprehensive metric.
    \textbf{(b) Precise Evidence Chains Generation:} The candidate chains are structured into a prefix tree. An LLM then performs a fine-grained, constrained decoding within this space to generate precise evidence chains.
    \textbf{(c) Retrieved Context Optimization:} The generated evidence chains are used to filter the retrieved documents, removing noise (e.g., $d_3$, $d_5$) and explicitly presenting the most relevant documents ($d_1$, $d_4$) along with the chains to the LLM.}
    \vspace{-1em}
    \label{fig:overall}
\end{figure*}
In this section, we introduce NeocorRAG, a novel framework to enhance model reasoning in QA tasks by addressing the dual needs of maintaining high recall and optimizing retrieval quality. First, we define the RCR to quantify the contribution of retrieved texts to reasoning and use this evaluation metric to guide retrieval optimization (Section~\ref{ssec:3.1}). Second, we propose a scheme to achieve this optimization by thoroughly mining deep evidence chains within the retrieved texts (Section~\ref{ssec:3.2}). Finally, we detail the implementation of evidence chains mining and retrieval optimization, which includes the identification of the candidate evidence space, the constrained generation of evidence chains, and the refinement guided by evidence chains of the retrieved contexts (Section~\ref{ssec:3.3}).

\subsection{Necessity of RCR for Retrieval Optimization}
\label{ssec:3.1}

Addressing the critical shortcoming of existing retrieval metrics, which overlook the actual contribution of retrieved texts to reasoning performance, we pioneer the \textbf{Recall Conversion Rate (RCR)} to quantify this contribution. Its core premise is that the value of retrieval lies not only in ``whether relevant information is retrieved'', but also, more importantly, in ``whether the retrieved information can be effectively converted into correct answers''. To unify existing evaluation metrics, RCR is defined as the ratio of the F1 score of the generated answer to the retrieval coverage, with the formula as follows:


\vspace{-1em}

\begin{equation}
\mathrm{RCR}(\mathcal{Q}, \mathcal{D}_g, A, A^*)=
\begin{cases}
\dfrac{\mathrm{F1}(A, A^*)}
{\mathrm{Recall}(\mathcal{D}_g,\mathcal{Q})},
& \text{if } \mathrm{Recall}(\mathcal{D}_g,\mathcal{Q})>0 \\
0, & \text{otherwise}
\end{cases}
\end{equation}

Here, $\mathcal{Q}$ denotes the set of queries, and $A$ and $A^*$ represent the generated answer and the ground-truth answer for each query, respectively. $\mathrm{F1}(A, A^*)$ measures the accuracy of the answer, while $\mathrm{Recall}(\mathcal{D}_g, \allowbreak \mathcal{Q})$ measures the retrieval coverage,
which can be instantiated as $\mathrm{Recall@}n$. Unless otherwise specified, we instantiate $\mathrm{Recall}$ as $\mathrm{Recall@5}$ in all experiments, and denote the resulting metric
as $\mathrm{RCR@5}$. Specifically, for each query $q\in\mathcal{Q}$, let $\mathcal{D}_g(q)$ denote its supporting passages; $\mathrm{Recall}@n$ is computed as the average fraction of $\mathcal{D}_g(q)$ that appears in the top-$n$ retrieved passages.


Based on the strong recall baseline of HippoRAG2, our approach improves retrieval efficacy by specifically targeting an improvement in RCR while minimizing degradation in recall performance. This allows our method to effectively navigate the fundamental trade-off between high recall and high utility in retrieved contexts.

\subsection{Evidence Chains Mining}
\label{ssec:3.2}

To optimize retrieval quality without sacrificing recall, it is essential to thoroughly mine the evidence information within the retrieved texts that supports reasoning. We argue that the reasoning value of the retrieved texts is not uniformly distributed but concentrated in chains that connect entities, relationships, and facts related to the query. Therefore, we define such structured sequences of key information in retrieved texts that support reasoning as ``evidence chain'' and have implemented a recall optimization algorithm based on these evidence chains.

The crux of this optimization lies in a faithful method for mining these evidence chains. We design a high-efficiency mining strategy characterized by ``broad candidate generation followed by strict filtering''. A novel activation-based path search algorithm ensures efficient exploration of the potential evidence space, while constrained decoding leverages the global representation power of generative models to precisely identify valid evidence chains within this space. The specific framework, illustrated in Figure~\ref{fig:overall}, consists of three main stages:
\begin{enumerate}[label=(\arabic*), itemsep=0.2em, topsep=0.2em, leftmargin=*]
    \item \textbf{Candidate Evidence Space Identification}: Lock the candidate evidence space within the subgraph constructed from the retrieved documents.
    \item \textbf{Precise Evidence Chains Generation}: Distill faithful and valid evidence chains from the candidate space.
    \item \textbf{Retrieved Context Optimization}: Filter out noisy texts and explicitly surface hidden relational evidence based on the set of evidence chains.
\end{enumerate}

\subsection{Evidence-Aware Retrieval Optimization}
\label{ssec:3.3}

\subsubsection{Candidate Evidence Chains Search}

This stage aims to mine potential evidence chains by conducting node identification and path exploration over the subgraph $G = (V, E)$ constructed from retrieved texts. The process progresses from initializing relevant nodes to searching and scoring candidate evidence chains, comprising the following steps:

\paragraph{Step 1: Initial Node Identification}

Given a query $q$, we extract a set of keywords entities $E_q = \{e_1, e_2, \ldots, e_m\}$ using an LLM. Based on these entities, we identify the relevant initial nodes by selecting those whose semantic embeddings are sufficiently close to the aggregated embeddings of $E_q$. Formally:
\begin{equation}
    V_{\text{init}} = \left\{ v \in V \,\middle|\, \phi_{\text{sim}}(\operatorname{Emb}(v), \operatorname{Emb}(E_q)) \geq \tau_{\text{node}} \right\}
\end{equation}
where $\operatorname{Emb}(\cdot)$ denotes a pretrained embedding model, $\phi_{\text{sim}}(\cdot, \cdot)$ is the semantic similarity function, and $\tau_{\text{node}} \in [0, 1]$ is the threshold for identifying query-relevant nodes.

\paragraph{Step 2: Dynamic Path Exploration}

To efficiently discover potential evidence chains, we employ a dynamic exploration algorithm based on Depth-First Search (DFS). Starting from the initial node set $V_{\text{init}}$, the algorithm recursively explores the paths under a maximum hop constraint $K$. At each exploration step, for a current node $v_i$ and its neighbor $n_j$ connected through the relation $r$, we form a triple $t = (v_i, r, n_j)$ and compute its local semantic similarity to the query:
\begin{equation}
    S(v_i, n_j) = \cos\left(\operatorname{Emb}(t), \operatorname{Emb}(q)\right)
\end{equation}
The search continues to expand only if the triple score satisfies the edge-level threshold condition. Specifically, low-confidence neighbors are pruned and only those satisfying the following constraint are included in the search frontier:
\begin{equation}
    n_j \in \text{Expanded Nodes} \iff S(v_i, n_j) \geq \tau_{\text{edge}}
\end{equation}

\paragraph{Step 3: Activated Comprehensive Path Scoring}

Local similarity scores alone are insufficient to evaluate the overall quality of an evidence chain. We therefore design a comprehensive scoring function that incorporates both penalties and rewards. Given a candidate path $P = [t_1, t_2, \ldots, t_{|P|}]$, the final score is calculated as:

\begin{equation}
\begin{split}
    \text{score}(P) =\ & 
    \underbrace{\alpha^{\text{ReLU}(|P| - L)}}_{\text{Length Penalty}} \cdot 
    \underbrace{\cos\left( \frac{1}{|P|} \sum_{k=1}^{|P|} \operatorname{Emb}(t_k),\, \operatorname{Emb}(q) \right)}_{\text{Overall Semantic Relevance}} \cdot \\
    & \underbrace{\beta^{\sum_{k=1}^{|P|} \mathbb{I}\left( \cos\left( \operatorname{Emb}(t_k), \,\operatorname{Emb}(q) \right) \geq \tau_{\text{boost}} \right)}}_{\text{High-Confidence Activation Reward}}
\end{split}
\end{equation}

The comprehensive scoring function balances three core components. The \textbf{length penalty} term $\alpha^{\text{ReLU}(|P| - L)}$ applies a penalty to paths whose length exceeds the expected maximum $L$, where $\alpha \in (0, 1)$ controls the penalty strength and $\text{ReLU}(x) = \max(0, x)$ ensures that no penalty is applied when the length is within the limit. The \textbf{overall semantic relevance} is measured by the cosine similarity between the average embedding of all triples in the path and the query embedding, capturing the global alignment between the candidate chain and the original question. We set $\tau_{\text{boost}} \ge \tau_{\text{edge}}$, so that only triples that pass the exploration threshold can further contribute to activation rewards. The \textbf{high-confidence activation reward} term uses an exponential form with base $\beta > 1$, rewarding paths that contain more triples whose semantic similarity to the query exceeds a predefined confidence threshold $\tau_{\text{boost}}$, thus boosting the score of chains enriched with highly relevant information. Here, $\mathbb{I}(\cdot)$ is an indicator function that returns 1 if the condition is true and 0 otherwise. 


\subsubsection{Precise Evidence Chains Generation}

To generate precise evidence chains from the candidate set obtained in the previous stage, we organize the candidate chains into a tree structure of prefixes~\cite{Triememory}. This tree-based organization facilitates constrained decoding for the LLM, ensuring that the next generated token is always a child of the current token's corresponding node in the prefix tree. This guaranties the precision of the evidence chains generated \citep{Knowledgegraphcompletion,Graph-constrained}.

Given the input query \(q\) and the set of retrieved documents \(\mathcal{T}\), we design an instruction prompt to guide the LLM's decoding process under the constraints imposed by the prefix tree \(\mathcal{T}_{\text{prefix}}\). This process effectively leverages the LLM's representational power to search within the candidate chain set, and can be formalized as:
\begin{equation}
\label{eq:faithful_generation}
\begin{multlined}
    P_{\theta}(\mathbf{e} | q, \mathcal{T}) = \underbrace{P_{\theta}(\mathbf{e} | q)}_{\text{Vanilla Decoding}} \\
    \cdot \underbrace{\prod_{i=1}^{|\mathbf{e}|} P_{\theta}(e_i | q, e_1, \ldots, e_{i-1}) \cdot C_{\mathcal{T}_{\text{prefix}}}(e_i | e_1, \ldots, e_{i-1})}_{\text{Evidence-Constrained Decoding}}
\end{multlined}
\end{equation}
where \(\theta\) represents the LLM parameters, \(\mathbf{e} = [e_1, e_2, \ldots, e_{|\mathbf{e}|}]\) denotes the generated evidence chain (a sequence of tokens), and \(C_{\mathcal{T}_{\text{prefix}}}(e_i | e_1, \ldots, e_{i-1})\) is a constraint function that verifies whether the partial sequence \([e_1, \ldots, e_i]\) is a valid prefix of any candidate chain in the prefix tree \(\mathcal{T}_{\text{prefix}}\):
\begin{equation}
\label{eq:constraint_function}
C_{\mathcal{T}_{\text{prefix}}}(e_i | e_1, \ldots, e_{i-1}) = 
\left\{
\begin{aligned}
    1 \quad & \text{if } \exists p \in \mathcal{P} \text{ such that }\\
            &  [e_1, \ldots, e_i] \text{ is a prefix of } p \\
    0 \quad & \text{otherwise}
\end{aligned}
\right.
\end{equation}

\begin{table*}[!ht]
\centering
\small

\caption{
Overall performance on classical QA benchmarks. We report passage recall@5 (R5), answer F1 score (F1), and RCR@5 (RCR). Best results are highlighted in\textbf{bold}, * denotes statistical significance (p < 0.05) against comparable baselines Trace.
}
\label{tab:main_results}

\vspace{-1.2em} 

\renewcommand{\arraystretch}{1.0}
\begin{tabular*}{\textwidth}{@{\extracolsep{\fill}}lccccccccccccccc@{}}
\toprule
\multirow{3}{*}{\textbf{Retrieval}} &
\multicolumn{6}{c}{\textbf{Simple QA}} &
\multicolumn{6}{c}{\textbf{Multi-Hop QA}} &
\multicolumn{3}{c}{\textbf{Avg.}} \\
\cmidrule(lr){2-7}\cmidrule(lr){8-13}\cmidrule(lr){14-16}
& \multicolumn{3}{c}{\textbf{NQ}} & \multicolumn{3}{c}{\textbf{MuSiQue}} &
  \multicolumn{3}{c}{\textbf{2Wiki}} & \multicolumn{3}{c}{\textbf{HotpotQA}} &
   &  &  \\
\cmidrule(lr){2-4}\cmidrule(lr){5-7}\cmidrule(lr){8-10}\cmidrule(lr){11-13}
& R5 & F1 & RCR & R5 & F1 & RCR &
  R5 & F1 & RCR & R5 & F1 & RCR &
  R5 & F1 & RCR \\
\midrule

\rowcolor{gray!20}
\multicolumn{16}{c}{\textbf{Llama-3.3-70B}} \\
\midrule
\multicolumn{16}{l}{\emph{Simple Baselines}} \\
BM25        & 56.1 & 59.0 & 105.2 & 43.5 & 28.8 & 66.2 & 65.3 & 51.2 & 78.4 & 74.8 & 62.3 & 83.3 & 59.9 & 50.3 & 83.3 \\
Contriever   & 54.6 & 58.9 & 107.9 & 46.6 & 31.3 & 67.2 & 57.5 & 41.9 & 72.9 & 75.3 & 63.4 & 84.2 & 58.5 & 48.9 & 83.1 \\
GTR (T5-base) & 63.4 & 59.9 & 94.5 & 49.1 & 34.6 & 70.5 & 67.9 & 52.8 & 77.8 & 73.9 & 62.8 & 85.0 & 63.6 & 52.5 & 82.0 \\
\midrule
\multicolumn{16}{l}{\emph{Structure-Enhanced RAG}} \\
RAPTOR       & 68.3 & 50.7 & 74.2 & 57.8 & 28.9 & 50.0 & 66.2 & 52.1 & 78.7 & 86.9 & 69.5 & 80.0 & 69.8 & 50.3 & 70.7 \\
HippoRAG     & 44.4 & 55.3 & 124.5 & 53.2 & 35.1 & 66.0 & 90.4 & 71.8 & 79.4 & 77.3 & 63.5 & 82.1 & 66.3 & 56.4 & 88.0 \\
HippoRAG2    & 78.0 & 63.3 & 81.2 & 74.7 & 48.6 & 65.1 & 90.4 & 71.0 & 78.5 & 96.3 & 75.5 & 78.4 & 84.9 & 64.6 & 75.8 \\
\midrule
\multicolumn{16}{l}{\emph{Reasoning-Enhanced RAG}} \\
CoRAG      & - & 54.5 & - & - & 52.9 & - & - & 75.1 & - & - & 75.1 & - & - & 64.4 & - \\
Trace (Llama-3.2-3B)      & 77.9 & 48.7 & 62.6 & 73.3 & 33.4 & 45.6 & 90.5 & 59.9 & 66.1 & 96.4 & 68.9 & 71.5 & 84.5 & 52.7 & 61.4 \\
Trace (Llama-3-8B)      & 77.9 & 50.1 & 64.3 & 73.3 & 39.1 & 53.4 & 90.5 & 65.4 & 72.2 & 96.4 & 73.1 & 75.9 & 84.5 & 56.9 & 66.4 \\
\midrule
\textbf{Ours (Llama-3.2-3B)} & 77.9 & 64.9* & 83.3 & 73.3 & 50.8* & 69.3 & 90.5 & 71.6* & 79.1 & 96.4 & 77.1* & 79.9 & 84.5 & 66.1 & 77.9 \\
\textbf{Ours (Llama-3-8B)} & 77.9 & \textbf{65.6*} & 84.3 & 73.3 & \textbf{52.6*} & 71.7 & 90.5 & \textbf{76.1*} & 84.1 & 96.4 & \textbf{78.3*} & 81.2 & 84.5 & 68.1 & 80.3 \\
\midrule

\rowcolor{gray!20}
\multicolumn{16}{c}{\textbf{Llama-3.2-3B}} \\
\midrule
HippoRAG2    & 78.0 & 49.9 & 63.9 & 74.7 & 30.4 & 40.7 & 90.4 & 46.6 & 51.5 & 96.3 & 55.7 & 57.8 & 84.9 & 45.7 & 53.5 \\
CoRAG     & - & 37.8 & - & - & 28.1 & - & - & 42.3 & - & - & 56.0 & - & - & 41.1 & - \\
Trace (Llama-3.2-3B)      & 77.9 & 47.5 & 61.0 & 73.3 & 24.6 & 33.5 & 90.5 & 40.1 & 44.3 & 96.4 & 61.9 & 64.2 & 84.5 & 43.5 & 50.8 \\
Trace (Llama-3-8B)     & 77.9 & 46.9 & 60.3 & 73.3 & 27.4 & 37.4 & 90.5 & 45.7 & 50.5 & 96.4 & 64.5 & 66.9 & 84.5 & 46.1 & 53.8 \\
\textbf{Ours (Llama-3.2-3B)} & 77.9 & 57.0* & 73.1 & 73.3 & 33.5* & 45.6 & 90.5 & 49.9* & 55.1 & 96.4 & 66.8* & 69.3 & 84.5 & 51.8 & 60.8 \\
\textbf{Ours (Llama-3-8B)} & 77.9 & \textbf{57.2*} & 73.5 & 73.3 & \textbf{34.8*} & 47.4 & 90.5 & \textbf{51.4*} & 56.8 & 96.4 & \textbf{67.0*} & 69.5 & 84.5 & 52.6 & 61.8 \\
\bottomrule
\end{tabular*}

\vspace{-0.8em}

\end{table*}

Here, \(\mathcal{P}\) is the set of candidate evidence chains. The constraint function \(C_{\mathcal{T}_{\text{prefix}}}\) ensures that all generated tokens correspond to nodes in the prefix tree \(\mathcal{T}_{\text{prefix}}\), thereby preventing hallucinations. This approach not only guarantees the precision of the generated chains but also bridges the gap between the LLM's parametric knowledge and the factual knowledge in the retrieved texts, helping to uncover hidden information.

In practice, answering a question often relies on multiple evidence chains, and different chains can help the model reason more comprehensively \citep{Self-consistency}. Therefore, we employ beam search~\cite{freitag2017beam} to obtain multiple evidence chains, which serve as supplementary evidence to the initially retrieved documents.
\subsubsection{Evidence-Guided Document Filtering}
\label{subsubsec:filtering}

In the final stage, we use the set of precise evidence chains to guide the refinement of the retrieved context. We first partition the initial document set \(D\) according to its retrieval ranking into a high-confidence set \(D_{\text{topN}}\), which is preserved by default, and a low-confidence set \(D_{\text{low}}\). The filtering process is then applied exclusively to \(D_{\text{low}}\). Specifically, we retain low-confidence documents that contain at least one triple from the union of all chains (\(\mathcal{T}_{\text{union}}\)), forming a supplementary set:
\begin{equation}
\label{eq:supplement_docs}
D_{\text{supplement}} = \left\{ d \in D_{\text{low}} \mid \exists t \in \mathcal{T}_{\text{union}}, t \subseteq d \right\}
\end{equation}
where \(t = (v_i, r, v_j)\) denotes a triple in an evidence chain, and \(t \subseteq d\) indicates that the linearized triple string appears as a substring of the document \(d\).
The final refined result is \(D_{\text{refined}} = D_{\text{topN}} \cup D_{\text{supplement}}\). This reduction in the total document count, observed in our experiments, occurs because the set of salvaged documents (\(D_{\text{supplement}}\)) is typically much smaller than the original low-confidence set (\(D_{\text{low}}\)). This targeted approach preserves core information while salvaging valuable evidence and eliminating substantial noise.

\section{Experiments}
\label{sec:experiments}

This section details our experimental setup, results, and analyzes, designed to answer the following four key research questions:
\begin{itemize}[leftmargin=*, itemsep=0pt, topsep=2pt]
    \item \textbf{RQ1:} Does \ours{} demonstrate superior efficacy as a retrieval optimization method over SOTA baselines?
    \item \textbf{RQ2:} Is \ours{} a low-cost, plug-and-play, and highly adaptable framework?
    \item \textbf{RQ3:} Are all constituent modules within the \ours{} framework indispensable for its overall performance?
    \item \textbf{RQ4:} How does the evidence chain mechanism in \ours{} align with and materialize the proposed retrieval quality standards?
\end{itemize}

\subsection{Experimental Setup}
\label{ssec:exp_setup}

\paragraph{\textbf{Datasets.}}
To comprehensively evaluate the performance of \ours{}, we select four QA datasets covering varying levels of complexity: \textbf{Simple QA:} Natural Questions (NQ)~\cite{kwiatkowski2019NQ}, used to test single-hop fact retrieval capabilities. \textbf{Multi-Hop QA:} MuSiQue~\citep{trivedi2022musique}, 2WikiMultiHopQA~\citep{2WikiMultiHopQA}, and HotpotQA~\citep{yang2018hotpotqa}. These datasets require the model to integrate multiple pieces of information to infer the answer, evaluating its associative and reasoning abilities. In particular, we use the same corpus and datasets as HippoRAG~\citep{jimenez2024hipporag}, with further details in Appendix~\ref{app:dataset_details}.

\paragraph{\textbf{Baselines.}}
To fully validate the performance advantages of \ours{}, we select three categories of retrieval methods as baselines: \textbf{Simple Baselines}, including BM25~\citep{robertson1995okapi}, Contriever~\citep{contriever}, and GTR~\citep{ni2021gtr}. \textbf{Structure-Enhanced RAG}, including RAPTOR~\citep{sarthi2024raptor}, HippoRAG~\citep{jimenez2024hipporag}, and HippoRAG2~\citep{gutierrez2025hipporag2}. \textbf{Reasoning-Enhanced RAG}, including CoRAG~\citep{wang2024corag} and Trace~\citep{fang2024trace}. To ensure a fair comparison, all baseline methods were evaluated under a unified experimental configuration to the greatest extent possible. See Appendix~\ref{app:implementation_details} for more details.

\paragraph{\textbf{Evaluation Metrics.}}
Consistent with previous studies~\citep{sarthi2024raptor, jimenez2024hipporag, gutierrez2025hipporag2}, we use the \textbf{Recall@5} and \textbf{F1 score} to evaluate retrieval capacity and reasoning performance, respectively. Additionally, to measure the efficiency of converting retrieval performance into reasoning effectiveness, we use the \textbf{RCR}.

\begin{figure}[t]
	\centering
	\includegraphics[width=\columnwidth]{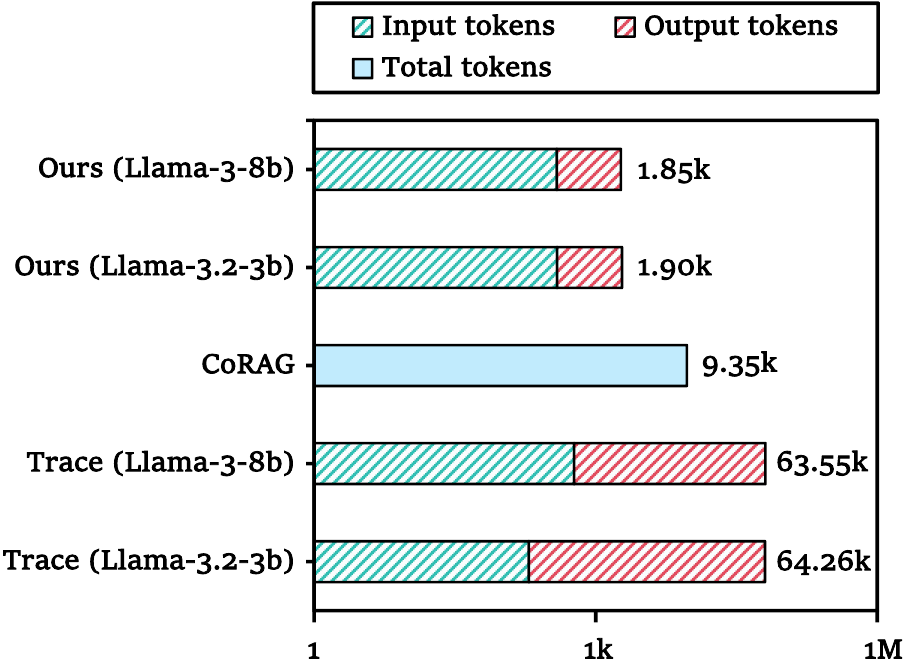} 

    \caption{Token efficiency comparison (tokens per question across four QA benchmarks).} 
	\label{fig:token}
\end{figure}
\begin{table}[!ht]
\centering
\small

\caption{
Ablation study results of NeocorRAG components.
}

\vspace{-1.2em} 

\renewcommand{\arraystretch}{1.1}
\newcommand{\upchange}[1]{\textcolor{red}{$\uparrow$#1}}
\newcommand{\downchange}[1]{\textcolor{green}{$\downarrow$#1}}
\begin{tabular*}{\linewidth}{@{\extracolsep{\fill}}lcccc}
\toprule
\textbf{Method} & \textbf{EM} & \textbf{F1} & \textbf{RCR@5} \\
\midrule
NeocorRAG       & 53.7 & 66.8 & 69.3 \\
w/ Naive Chain Search  & 35.8 ($\downarrow$17.9) & 49.5 ($\downarrow$17.3) & 51.3 ($\downarrow$18.0) \\
w/o Constrained Decoding    & 45.1 ($\downarrow$8.6) & 58.7 ($\downarrow$8.1) & 61.0 ($\downarrow$8.3) \\
w/o Document Filtering & 43.3 ($\downarrow$10.4) & 57.0 ($\downarrow$9.8) & 59.1 ($\downarrow$10.2) \\
\bottomrule
\end{tabular*}

\vspace{-1.8em} 
\label{tab:ablation3}
\end{table}

\paragraph{\textbf{Implementation Details.}}
\ours{} is based on the retrieval results of HippoRAG2~\citep{gutierrez2025hipporag2}, using the same initial retrieval configuration. For our core retrieval quality optimization stage, we employ lightweight models from the Llama~\cite{dubey2024llama3herd}, Qwen~\cite{qwen2.5} series as the constrained decoding model for mining evidence chains and the bge-large-en-v1.5 model~\citep{BGE,bge_embedding} as the general-purpose encoding model. To rigorously assess the generalizability and model-scale invariance of retrieval quality enhancements, our evaluation employs a dual-model framework utilizing both Llama-3.3-70B-Instruct (large-scale) and Llama-3.2-3B-Instruct (small-scale) as downstream answer generators. All experiments were conducted on 8 NVIDIA A40 GPUs (48GB VRAM each). We report the average results over 5 independent runs with a fixed random seed of 42. See Appendix~\ref{app:implementation_details} for more details. 
\subsection{Main Results (RQ1)}
\label{ssec:main_results}

As shown in Table~\ref{tab:main_results}, \ours{} comprehensively outperforms all baselines on all benchmarks and model sizes (3B and 70B), validating its superior effectiveness in QA tasks.

Compared to HippoRAG2~\cite{gutierrez2025hipporag2}, with identical recall rates and nearly identical QA prompts, \ours{} achieves average improvement of the F1 score of 6.9\% and 3.5\%, and average RCR improvements of 8.3\% and 4.5\% in the 3B and 70B settings, respectively. On HotpotQA, \ours{} achieves an improvement in F1 of up to 9.4\% and an increase in RCR of 9.6\%. This performance confirms that optimizing retrieval quality enhances reasoning performance.


Furthermore, \ours{} exceeds the reasoning-enhanced baselines. The average performance gap in the F1 score against Trace \cite{fang2024trace} increases from 6.5\% on 3B models to 11.2\% on 70B models. This is because its shallow, title-based evidence tracking can erroneously discard correct documents. This strategy becomes counterproductive with more powerful foundation models. Similarly, CoRAG \cite{wang2024corag} underperforms in the 3B model by 11.5\% in the F1 score, as its dynamic retrieval relies too heavily on the models' capabilities. The significant performance gaps on both multi-hop and simple question answering also indicate its poor generalization ability. In contrast, \ours{} follows a more systematic guideline to optimize retrieval quality by deeply mining hidden evidence chains. This approach ensures greater comprehensiveness while reducing information noise and enabling evidence visualization, leading to stable and significant performance gains.
\vspace{-3pt}

\subsection{Applicability and Efficiency (RQ2)}
\label{ssec:applicability}
\textbf{Adaptability to Different Base LLMs.}
To thoroughly validate the adaptability of \ours{}, we conducted experiments using two mainstream open-source model families for evidence chain mining: Llama 3 and Qwen2.5, with parameter sizes ranging from 1B to 14B.
Across different model families and scales, \ours{} exhibits stable performance with only minor variations, indicating that its effectiveness is not tied to a specific backbone model.
In particular, for models below the 3B parameters, increasing the number of candidate evidence chains may lead to performance degradation due to limited model capacity, whereas models beyond 3B consistently benefit from expanded candidate evidence chains.
Additional diagnostic results and detailed comparisons between model families and evidence chain counts are provided in Appendix~\ref{app:fig6}.
These findings indicate that after Candidate Evidence Space Identification, the refined candidate space requires only lightweight models for Precise Evidence Chain Generation, eliminating the need for excessively powerful models.

\textbf{Token Cost Analysis.}
\ours{} significantly reduces token consumption on all benchmarks compared to reasoning-enhanced baselines. On average, our method uses only 20.1\% of the tokens required by CoRAG~\cite{wang2024corag} and less than 2.94\% of the tokens used by Trace~\cite{fang2024trace}. This high efficiency is attributed to our effective evidence chain mining process, as further illustrated in Figure~\ref{fig:token}.

\textbf{Summary.} 
In summary, \ours{} is adaptable to various model families and scalable from 1B to 14B parameters, maintaining stable performance without complex adjustments. Its design, centered on lightweight auxiliary models, ensures both efficiency and low computational cost. Together, these characteristics answer RQ2: \ours{} is a broadly applicable and highly efficient framework with practical value. We further report runtime latency and GPU memory overhead on HotpotQA in Appendix~\ref{app:efficiency}.

\begin{figure}[t]
	\centering
    \includegraphics[width=\columnwidth]{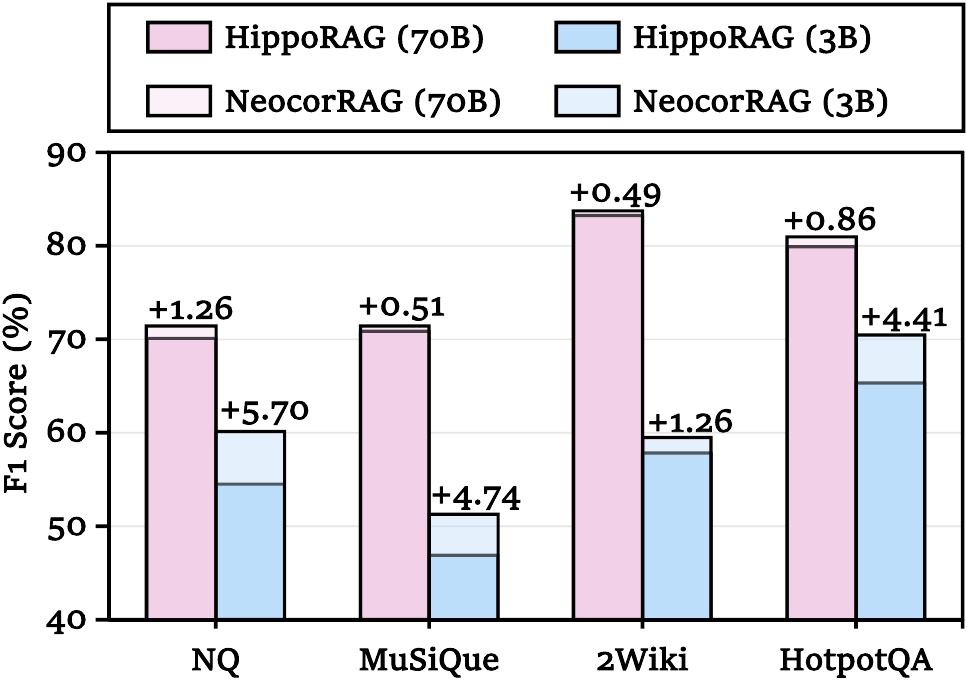} 
    \caption{Contribution of evidence chains to the F1 score when using ground truth (golden) documents. This setup isolates the benefit of explicating hidden information from the task of filtering irrelevant documents.} 

    \vspace{-2.5em} 
	\label{fig:godlen}
\end{figure}
\subsection{Ablation Study (RQ3)}
\label{ssec:ablation_study}

To answer RQ3, we conducted a series of ablation studies on the HotpotQA benchmark using the Llama-3.2-3B-Instruct model. The results presented in Table~\ref{tab:ablation3}, evaluate the effectiveness of each core component of \ours{}. Our key observations are as follows:
\begin{enumerate}[label=(\arabic*), itemsep=0.2em, topsep=0.2em, leftmargin=*]
    \item Removal of any single component leads to a significant performance drop, confirming the necessity of each part of our design.
    \item Replacing our activated search with a naive chain search results in the most substantial performance degradation, even below the variant without document filtering. This underscores the critical importance of a high-quality candidate space for the subsequent generation of evidence. Additional robustness and sensitivity diagnostics under variations in candidate evidence space construction are reported in Appendix~\ref{app:robustness_sensitivity}.
    \item Removal of the prefix tree constraints allows the generation of numerous unfaithful chains from the refined candidate set. Using these for optimization still causes a significant decline in performance, highlighting the necessity of constrained decoding.
    \item Forgoing document filtering allows a large amount of noise to interfere with the model's reasoning. The resulting performance drop demonstrates the importance of actively removing this interference.
\end{enumerate}

\subsection{Effectiveness Validation of Evidence Chains against Retrieval Quality Standards (RQ4)}
\label{ssec:analysis_hidden_info}

To evaluate the capability of evidence chains to discover latent evidence and verify the role of evidence visualization in improving reasoning performance, we conducted experiments using ground truth documents with 3B and 70B parameter models. The results in Figure~\ref{fig:godlen} show that even in a completely noise-free environment, the evidence chains still improve the average performance of the 3B model by 4.03\%.  In contrast, the 70B model achieves only a limited gain of 0.78\% in noisy scenarios. These findings demonstrate that \ours{} effectively uncovers hidden evidence.  Moreover, explicit evidence is particularly critical for smaller-scale models and remains highly valuable in the presence of increased noise.

\begin{table}[!ht]
\centering
\small
\caption{
Impact of evidence chains on filtering redundant documents on four benchmarks. We report Average Total Documents (ATD) and Average Irrelevant Documents (AID) before and after applying evidence chain filtering. 
Last row shows the average change across benchmarks, where $\uparrow$ and $\downarrow$ indicate performance increases and decreases, respectively.
}
\label{tab:ablation2}

\renewcommand{\arraystretch}{1.1}
\newcommand{\upchange}[1]{\textcolor{red}{$\uparrow$#1}}
\newcommand{\downchange}[1]{\textcolor{green}{$\downarrow$#1}}

\vspace{-0.8em} 
\begin{tabular*}{\linewidth}{@{\extracolsep{\fill}}lcccc}
\toprule
\textbf{Filter} & \textbf{Recall@5} & \textbf{ATD} & \textbf{AID} & \textbf{RCR@5} \\
\midrule
\multicolumn{5}{l}{\emph{Before}} \\
NQ       & 77.90 & 5.00 & 2.33 & 63.90 \\
MuSiQue  & 73.30 & 5.00 & 3.12 & 40.70 \\
2Wiki    & 90.50 & 5.00 & 2.80 & 51.50 \\
HotpotQA & 96.40 & 5.00 & 3.08 & 57.80 \\
\textbf{Avg.}      & \textbf{84.53} & \textbf{5.00} & \textbf{2.83} & \textbf{53.48} \\
\midrule
\multicolumn{5}{l}{\emph{After}} \\
NQ       & 68.56 & 3.76 & 1.35 & 73.47 \\
MuSiQue  & 70.33 & 3.61 & 1.81 & 47.48 \\
2Wiki    & 89.10 & 3.44 & 1.28 & 56.78 \\
HotpotQA & 94.45 & 3.28 & 1.13 & 69.51 \\
\textbf{Avg.}      & \textbf{80.61} & \textbf{3.52} & \textbf{1.39} & \textbf{61.81} \\
\midrule
\multicolumn{5}{l}{\emph{Change}} \\
\textbf{Avg.} & $\downarrow$3.92 & $\downarrow$\textbf{1.48} & $\downarrow$\textbf{1.44} & $\uparrow$\textbf{8.33}\\
\bottomrule
\end{tabular*}

\vspace{-1.4em} 
\end{table}

Furthermore, we conducted an ablation study on the 3B model to verify the effectiveness of the evidence chain in filtering irrelevant documents. We introduce two metrics to quantify the filtering process's efficacy: \textbf{Average Total Documents (ATD)} and \textbf{Average Invalid Documents (AID)}. Detailed definitions and formulas for these metrics are provided in Appendix~\ref{app:metric_definitions}.
    

As shown in Table~\ref{tab:ablation2}, evidence-chain-based filtering reduces ATD by 1.48 and AID by 1.44 on average. This filtering procedure successfully filtered over half of the irrelevant content. Critically, this filtering process only slightly decreased the recall rate to 80.61\%. This figure is substantially higher than the 46.85\% recall achieved by the comparable Trace~\cite{fang2024trace} filtering method. These experimental results demonstrate that our method effectively enhances retrieval quality while safeguarding recall performance. This further validates the effectiveness of our proposed criteria for retrieval quality optimization.
\section{Conclusion}
\label{sec:conclusion}

This paper identifies limitations in existing retrieval methods through the lens of retrieval quality optimization. We establish finer-grained standards for high-quality retrieval and chart a new direction of improvement. To address current challenges, we propose an evidence-chain-based algorithm that refines retrieved texts to better support generative models' reasoning needs. By thoroughly mining evidence chains, we significantly reduce retrieval noise while explicitly exposing hidden evidence without recall loss, substantially improving reasoning performance. The SOTA results of \ours{} on four QA benchmarks validate our retrieval optimization approach.

\begin{acks}
This work is supported by the National Natural Science Foundation of China (Grant No. 62473271), and the Fundamental Research Funds for the Beijing University of Posts and Telecommunications (Grant No. 2025AI4S03). This work is also supported by the Engineering Research Center of Information Networks, Ministry of Education, China. We would like to thank Yuanze Li for assistance with figure design
and visualization in this paper. We would also like to thank the anonymous reviewers and area chairs for constructive discussions and feedback.
\end{acks}

\clearpage

\bibliographystyle{ACM-Reference-Format}
\balance
\bibliography{main}

\appendix

\section{Dataset Details}
\label{app:dataset_details}
In our experiments with \textbf{NeocorRAG}, we adopt the same Wikipedia snapshot and retrieval pipeline as \textbf{HippoRAG2}. Each sampled question is matched against the identical document corpus using consistent retrieval parameters, token limits, and filtering heuristics. This strict alignment ensures that any observed performance differences can be attributed to the reasoning capability of the model rather than retrieval variations or corpus drift.

\begin{table}[h!]
    \centering
    \caption{Detailed statistics of QA datasets used in our experiments. All samples are drawn from the hard subset to promote reasoning-intensive evaluation.}
    \label{tab:dataset_detail}
    \begin{tabular}{lcc}
        \toprule
        \textbf{Dataset} & \textbf{QA Type} & \textbf{Samples Used} \\
        \midrule
        HotpotQA              & Multi-hop     & 1,000 (hard subset) \\
        2WikiMultiHopQA       & Multi-hop     & 1,000 (hard subset) \\
        MuSiQue               & Multi-hop     & 1,000 (hard subset) \\
        NaturalQuestions (NQ) & Simple (Open) & 1,000 (nq-rear) \\
        \bottomrule
    \end{tabular}
    \vspace{-1em}
\end{table}

\section{Implementation Details}
\label{app:implementation_details}
\begin{table*}[!t]
\centering
\footnotesize
\caption{Hyperparameter settings of NeocorRAG.}
\label{tab:neocorrag_params}
\begin{tabular*}{\textwidth}{@{\extracolsep{\fill}}ll*{4}{c}}
\toprule
\textbf{Stage} & \textbf{Parameter} & \textbf{NQ} & \textbf{MuSiQue} & \textbf{2Wiki} & \textbf{HotpotQA} \\
\midrule
\multirow{5}{*}{\makecell[l]{\textbf{Candidate} \\ \textbf{Evidence Space} \\ \textbf{Identifcation}}}
& Node Relevance Threshold ($\tau_{\text{node}}$) & 0.90 & 0.90 & 0.90 & 0.90 \\
& Max Path Length ($L$) & 10 & 10 & 10 & 10 \\
& Activation Coefficient ($\beta$) & 1.10 & 1.10 & 1.10 & 1.10 \\
& Attenuation Coefficient ($\alpha$) & 0.90 & 0.90 & 0.90 & 0.90 \\
& Pruning Score Threshold ($\tau_{\text{edge}}$) & 0.45 & 0.45 & 0.45 & 0.45 \\
\midrule
\multirow{4}{*}{\makecell[l]{\textbf{Precise} \\ \textbf{Evidence Chains} \\ \textbf{Generation}}}
& Search Strategy & Beam & Beam & Beam & Beam \\
& Beam Width ($k$) & 3 & 5 & 5 & 5 \\
& Number of GCD Docs & 5 & 5 & 5 & 5 \\
& Number of Candidate Evidence Chains & 60 & 60 & 60 & 60 \\
\midrule
\makecell[l]{\textbf{Retrieved} \\ \textbf{Context} \\ \textbf{Optimization}}
& Reflection Top-N & 2 & 2 & 2 & 2 \\
\bottomrule
\end{tabular*}
\vspace{-1.5em}
\end{table*}
\subsection{Configuration of Other Baseline Methods}
We provide detailed implementation configurations for all baseline methods, with particular attention to \textbf{CoRAG} and \textbf{Trace}. For a fair comparison, we standardized the \textbf{retriever encoder model} across these methods to \textbf{e5-large-v2}. This unification guaranties a consistent foundation for retrieval input, enabling a more objective evaluation of the effectiveness of their respective \textbf{Reasoning-Chain-Enhanced RAG} frameworks.

For \textbf{Trace}, similar to our \textbf{NeocorRAG}, we utilized the retrieval results provided by the \textbf{HippoRAG} framework as input. This setup allows for a direct comparison of context optimization performance starting from the same initial retrieved document set. In contrast, \textbf{CoRAG} employs its own retrieval module.

\subsection{Configuration of Our Method}
Here, we detail the parameter settings used to optimize retrieved texts in \textbf{NeocorRAG}. Our central objective is to efficiently and comprehensively extract faithful evidence chains to guide retrieved text optimization.

During the final \textbf{Question Answering (QA)} stage, we adopted distinct prompting strategies depending on the scale of the employed Large Language Model (LLM):

\begin{itemize}[leftmargin=*]
\item When using a \textbf{3B-scale LLM}, \textbf{NeocorRAG} employed the \textbf{same prompt template as Trace} to ensure consistency during the QA phase with comparable model sizes.
\item When using a \textbf{70B-scale LLM}, \textbf{NeocorRAG} adopted the \textbf{same QA-stage prompt template as HippoRAG2}. This alignment ensures comparability with a strong RAG baseline using the same retrieval input and a high-capacity model.
\end{itemize}

Table~\ref{tab:neocorrag_params} presents the hyperparameter settings of the \textbf{NeocorRAG} framework on four benchmarks: NQ, MuSiQue, 2WikiMultiHopQA, and HotpotQA. These parameters span three key processing stages—\textit{Candidate Evidence Space Identification}, \textit{Precise Evidence Chain Generation}, and \textit{Retrieved Context Optimization}—each contributing to the effective enhancement of retrieved text quality.


\nobalance
\section{LLM Prompts}
\label{app:llm_prompts}

\begin{orangeanswer}\small
    \textbf{Context Passages:}

    \vspace{4pt}
    \noindent
    \textbf{• Wikipedia Title: The Newcomers (film)} \\
    The Newcomers is a 2000 American family drama film directed by James Allen Bradley and starring Christopher McCoy, Kate Bosworth, Paul Dano and Chris Evans. Christopher McCoy plays Sam Docherty, a boy who moves to Vermont with his family, hoping to make a fresh start away from the city. It was filmed in Vermont, and released by Artist View Entertainment and MTI Home Video.

    \vspace{6pt}
    \noindent
    \textbf{• Wikipedia Title: Chris Evans (actor)} \\
    Christopher Robert Evans (born June 13, 1981) is an American actor and filmmaker. Evans is known for his superhero roles as the Marvel Comics characters Steve Rogers / Captain America in the Marvel Cinematic Universe and Johnny Storm / Human Torch in "Fantastic Four" and .

    \vspace{8pt} 

    \parbox{\linewidth}{
        \ttfamily 
        \setlength{\parindent}{0pt} 
        \# Question: what is one of the stars of The Newcomers known for? \\[4pt]
        \# Information association path: ['The Newcomers - stars -> Chris Evans - known for role -> Steve Rogers / Captain America - part of -> Marvel Cinematic Universe - based on -> Marvel Comics', 'The Newcomers - stars -> Chris Evans - known for role -> Johnny Storm / Human Torch - part of -> Fantastic Four', 'The Newcomers - stars -> Chris Evans - profession -> actor']
    }

    \vspace{6pt} 

    \noindent
    \textbf{Thought:} Chris Evans, a star in The Newcomers, is known for superhero roles as Marvel Comics characters like Captain America and Human Torch.

    \vspace{4pt}

    \noindent
    \textbf{Answer:} superhero roles as the Marvel Comics.
\end{orangeanswer}

The following is the prompt template used for the 70B-scale LLM:
\begin{correctanswer}
\small
As an advanced reading comprehension assistant, your task is to analyze text passages and corresponding questions meticulously. \textbf{Note!:} Information association paths serve as reference pointers to surface latent connections. Your response start after "Thought:", where you will methodically break down the reasoning process, illustrating how you arrive at conclusions. Conclude with "Answer:" to present a concise, definitive response, devoid of additional elaborations.
\end{correctanswer}

The following is the prompt template used for the 3B-scale LLM:

\begin{correctanswer}
\small
Given some contexts, candidate evidence chains, and a question, the chain be used to help find potential associative information. please only output the answer to the question.
\end{correctanswer}

\begin{analyzeanswer}\small
    \vspace{2pt}
    {\ttfamily
    Input: \\
    Wikipedia Title: \{\} \\
    Question: \{\} \\
    Information association paths: \{\} \\
    Thought:\{\}
    }
    \vspace{2pt}
\end{analyzeanswer}

\section{Backbone Model Adaptability}
\label{app:fig6}
\begin{figure*}[t]
        \centering	
        \includegraphics[width=1\textwidth]{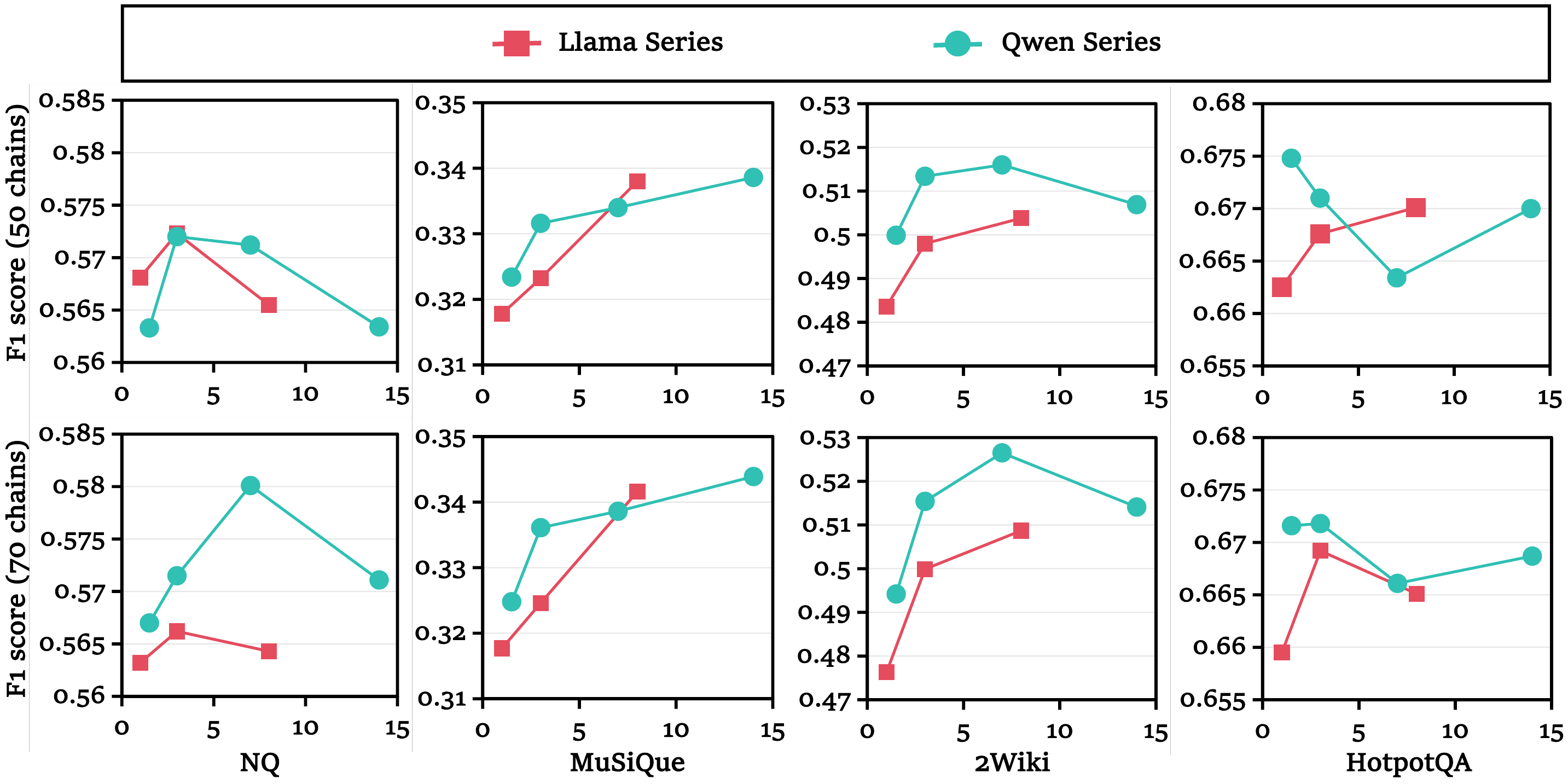}	\caption{F1 Score vs. evidence chain count across model families and sizes. This figure uses Llama-3.2-3B as the QA model and presents the performance comparison under two retrieval constraints (50 and 70 candidate evidence chains) on four QA benchmarks.}
\label{fig:6}
\vspace{-1.5em}
\end{figure*}

This appendix provides an additional diagnostic analysis on the adaptability of \ours{} to different base language models and evidence chain counts, supplementing the discussion in Section~\ref{ssec:applicability}.
Figure~\ref{fig:6} reports the performance of F1 between the model families (Llama and Qwen) and the model sizes under different constraints in the evidence chain of the candidate evidence.
The results further confirm that \ours{} maintains stable performance across the backbone models and that lightweight models benefit from carefully controlled evidence chain counts, while larger models can effectively exploit expanded candidate spaces.
These observations support our choice of design to use lightweight auxiliary models for the generation of evidence chains.

\section{Efficiency and Overhead}
\label{app:efficiency}
\textbf{Scope.}
In the main paper, we evaluate the efficiency of NeocorRAG primarily
through token cost analysis on four benchmarks,
showing that it uses only 20.1\% of the tokens required by CoRAG
and less than 2.94\% of those used by TRACE.
Although token consumption is a strong proxy for computational cost,
it does not fully capture practical efficiency.
We therefore further characterize runtime latency and GPU memory overhead.

\textbf{Experimental Setup.}
Following the protocol in Section~\ref{ssec:exp_setup},
we perform the efficiency analysis on \textbf{HotpotQA},
our most challenging multi-hop benchmark.
All experiments adopt identical retrieval results from HippoRAG2
and the same decoding configuration.
We use \textbf{Llama-3.2-3B-Instruct} as the constrained-decoding model
for the generation of precise evidence chain and the final QA.
Results are averaged over \textbf{5 runs} with a fixed random seed of \textbf{42}.

\textbf{Latency and Memory.}
In single-instance QA, NeocorRAG achieves a \textbf{speedup of 3.73$\times$} 
over TRACE and a \textbf{speedup of 1.27$\times$} over CoRAG,
consistent with token cost trends.
The maximum usage of GPU memory during pathfinding and inference is \textbf{10,380 MiB}.
Our constrained decoding further reduces peak memory by \textbf{13.7\%}
compared to a standard vLLM launch.

\textbf{Graph Overhead.}
The graph is constructed once during corpus ingestion.
At query time, required subgraphs are retrieved via local index lookups,
which introduces no additional GPU memory overhead.
In general, NeocorRAG achieves high efficiency and low computational cost.

\section{Robustness and Sensitivity}
\label{app:robustness_sensitivity}

\textbf{Scope.}
We provide additional diagnostics to examine the robustness and sensitivity
of NeocorRAG under variations in graph construction and hyperparameters.
All experiments are conducted on \textbf{HotpotQA},
using the same protocol, random seed and five-run averaging as in Section~\ref{ssec:exp_setup}.
We use \textbf{Llama-3.2-3B} for evidence chain mining
and report changes in F1 score.

\textbf{Robustness to Candidate Evidence Space Variations.}
We evaluate robustness to variations in triple quality,
graph expansion depth, and retriever encoders.
Injecting 50\% noise into triples decreases F1 by \(\approx\)0.43 points,
while randomly dropping 10\% increases F1 by \(\approx\)0.8 points.
Setting \texttt{max\_path\_length}=7 decreases F1 by \(\approx\)0.43,
whereas expanding to 13 improves F1 by \(\approx\)0.8.
Replacing \texttt{bge-large} with \texttt{bge-base} or \texttt{bge-small}
reduces F1 by \(\approx\)0.39 and \(\approx\)0.49 points, respectively.
In general, NeocorRAG exhibits predictable performance variations
without brittle behavior.

\textbf{Sensitivity to Evidence Chain Generation Hyperparameters.}
We further study the sensitivity of NeocorRAG to key hyperparameters in
the evidence chain generation stage.
Varying the number of candidate evidence chains from 40 to 70 results in
F1 changes of less than 0.2 percentage points, indicating low sensitivity
to this hyperparameter and demonstrating that NeocorRAG does not rely on
narrowly tuned parameter choices to achieve strong performance.

\section{Metric Definitions}
\label{app:metric_definitions}
This section provides detailed definitions of the metrics used to quantify the efficacy of the filtering process in our ablation study (Section~\ref{ssec:analysis_hidden_info}).

\begin{itemize}[leftmargin=*, itemsep=0pt, topsep=2pt]
    \item \textbf{Average Total Documents (ATD)}: The average number of documents provided to the reasoning model:
    \begin{equation}
    \text{ATD} = \frac{1}{n} \sum_{i=1}^{n} \left| D_i' \right|
    \end{equation}
    where \( D_i' \) denotes the final set of documents used by the model for the $i$-th question, and \( n \) is the total number of questions.
    
    \item \textbf{Average Invalid Documents (AID)}: The average number of irrelevant documents among those provided:
    \begin{equation}
    \text{AID} = \frac{1}{n} \sum_{i=1}^{n} \left| \{ d \mid d \in D_i', d \notin D_i^c \} \right|
    \end{equation}
    where \( D_i^c \) denotes the ground truth document set for the $i$-th question.
\end{itemize} 

\end{document}